 \documentclass[final,5p,times,twocolumn,authoryear]{elsarticle}

\usepackage{amssymb}
\usepackage{lipsum}

\usepackage{ulem}
\usepackage[dvipsnames]{color}

\def\squareforqed{\hbox{\rlap{$\sqcap$}$\sqcup$}}

\def\sq{\ifmmode\squareforqed\else{\unskip\nobreak\hfil
\penalty50\hskip1em\null\nobreak\hfil\squareforqed
\parfillskip=0pt\finalhyphendemerits=0\endgraf}\fi}

\def\utw{\smash{\rlap{\lower5pt\hbox{$\sim$}}}}

\def\udtw{\smash{\rlap{\lower6pt\hbox{$\approx$}}}}

\def\diameter{{\ifmmode\mathchoice
{\ooalign{\hfil\hbox{$\displaystyle/$}\hfil\crcr
{\hbox{$\displaystyle\mathchar"20D$}}}}
{\ooalign{\hfil\hbox{$\textstyle/$}\hfil\crcr
{\hbox{$\textstyle\mathchar"20D$}}}}
{\ooalign{\hfil\hbox{$\scriptstyle/$}\hfil\crcr
{\hbox{$\scriptstyle\mathchar"20D$}}}}
{\ooalign{\hfil\hbox{$\scriptscriptstyle/$}\hfil\crcr
{\hbox{$\scriptscriptstyle\mathchar"20D$}}}}
\else{\ooalign{\hfil/\hfil\crcr\mathhexbox20D}}%
\fi}}

\def\be{\begin{equation}}
\def\ee{\end{equation}}
\def\ba{\begin{eqnarray}}
\def\ea{\end{eqnarray}}

\def\msun{M_\odot}

\def\ltsima{$\; \buildrel < \over \sim \;$}
\def\simlt{\lower.5ex\hbox{\ltsima}}
\def\gtsima{$\; \buildrel > \over \sim \;$}
\def\simgt{\lower.5ex\hbox{\gtsima}}

\definecolor{webgreen}{rgb}{0,.5,0}
\definecolor{webbrown}{rgb}{.6,0,0}
\definecolor{falured}{rgb}{0.5, 0.09, 0.09}

\journal{New Astronomy}

\begin{document}

\begin{frontmatter}



\title{Inhibited destruction of dust by supernova in a clumpy medium}

\author[]{Svyatoslav Yu. Dedikov}
\ead{s.dedikov@asc.rssi.ru}
\author[]{Evgenii O. Vasiliev}
\ead{eugstar@mail.ru}
 \address{{{Lebedev Physical Institute of Russian Academy of Sciences}, {53 Leninskiy Ave.}, {Moscow}, {119991}, {Russia}}}     
        


\begin{abstract}
The decrease rate of dust mass due to strong shock waves ($v_s\geq 150$ km s$^{-1}$) from supernovae (SNe) estimated for the Milky Way interstellar medium significantly exceeds the overall production rate by both asymptotic giant branch stars and core collapse SNe. The interplay between the production and destruction rates is critically important for evaluation of the net dust outcome from SNe at different conditions. In light of this, we study the dynamics of initially polydisperse dust grains { pre-existing} in an ambient medium swept up the SN shock front depending on magnitude of inhomogeneity (clumpiness) in {the} medium. We find that dust destruction inside the bubble is inhibited in more inhomogeneous medium: the larger amount of dust survives for the higher dispersion of density. This trend is set by the interrelation between radiative gas cooling and dust sputtering in different environment. After several radiative times the mass fraction of the survived dust saturates at the level almost independent on the gas mean density. We note that for more clumpy medium the distributions of dust over thermal phases of a gas inside the bubble and over sizes are  smoother and flatter in comparison with those in a nearly homogeneous medium.
\end{abstract}




\end{frontmatter}




\section{Introduction}
\label{introduction}

Over the past three decades several attempts to estimate the dust budget have revealed a divergence of such an inventory \citep[see for recent discussion in][]{Mattsson2021,Kirchschlager2022,Peroux2023}. The total dust production rate in the Milky Way interstellar medium (ISM) from planetary nebulae, red giant and supergiants, carbon star winds and supernovae II (SNII) are thought to supply $\dot M_d^+\sim 0.005~\msun$ yr$^{-1}$, when $\sim 0.1 \msun$ per SN is assumed \citep[see in ][]{Draine2009}. The growth of dust particles through coagulation in the ISM is also thought to be an efficient mechanism in denser ISM environment with enhanced accretion. However, recent considerations have demonstrated that this mechanism is rather sensitive to environmental conditions, and cannot be a reliable source of dust replenishment \citep{Ferrara2016,Ceccarelli2018,Priestley2021a}. 

It is generally believed that dust particles are efficiently { destroyed} behind strong shocks ($v_s\geq 150$ km s$^{-1}$) and in the hot gas ($T\simgt 10^6$~K) due to the inertial and thermal sputtering \citep[][ and references therein]{Barlow1978,Draine1979a,Draine1979b}, and the shattering in grain-grain collisions at higher densities \citep[][]{Borkowski1995,Jones1996,Bocchio2016}. The characteristic dust lifetime in the Milky Way ISM against sputtering is estimated { from}  $t_{sp}\simlt 3\times 10^8$ yr \citep[][]{McKee1989,Jones1994} to $t_{sp}\simlt 3\times 10^9$ yr \citep[][]{Jones1994b,Slavin2015}, resulting in the decrease rate $\dot M_d^-\simlt (0.1-0.01)\msun$ yr$^{-1}$; more recent discussion can be found in \citep[][]{Bocchio2014,Ginolfi2018,Micelotta2018,Ferrara2021}. 

Thus, there is a challenging disbalance between the dust destruction and its replenishment \citep[][]{Draine2009}. Several mechanisms can reduce this discrepancy, e.g. due to possible dust-to-gas decoupling \citep[][]{Hopkins2016,Mattsson2019,Mattsson2022}, growth of grains in the ISM \citep[][]{Draine1990,Chokshi1993,Dwek1998,Calura2008,Draine2009,Mattsson2011,Inoue2011,Ginolfi2018,Heck2020} and formation of dust in supersonic turbulence \citep{Hopkins2016,Mattsson2019,Mattsson2019b,Mattsson2020,Mattsson2020b,Li2020,Commercon2023}.
In general, such a strong difference between the $\dot M_d^-$ and the $\dot M_d^+$ requires any possibility for reducing the interstellar ({ pre-existing}) dust destruction rate behind shocks waves.

In this paper, we estimate the dust destruction efficiency of a SN remnant embedded into a clumpy medium. The dust sputtering rate is known to strongly decrease with temperature $\propto T^3$ below $T\sim 10^6$ K \citep{Draine-book}. In a clumpy medium a fraction of the shock { wave area which enters a dense clump, propagates inward it with the velocity} $v_s\propto \rho_c^{-1/2}$ and gas temperature $T\propto \rho_c^{-1}$. For typical overdensity in interstellar clouds and clumps $\delta=\rho_c/\rho_i\geq 10$ even { stronger shocks} with $v_s>150$ km s$^{-1}$ fall below the destruction threshold 50 km s$^{-1}$. Therefore, for a sufficiently large total area of clumps and cloudlets the overall destructive capacity from SNe shocks can be diminished considerably. 

In order to { quantitatively characterize the impact} of a ISM clumpiness on to dust destruction, we { develop} a 3D multi-fluid hydrodynamical { model} with polydisperse dust particles. The paper is structured as follows. In Section 2 we describe the details of the model. In Section 3 we present the dynamics of dust grains in an inhomogeneous medium. Section 4 contains a general discussion and summarizes the results.

\section{Model} 
\label{sec-m}

We consider dynamics and destruction of polydisperse dust particles in a supernova remnant over its {100 kyr-long} expansion in an inhomogeneous ISM. To { describe} the clumpy gas density field we use the module pyFC\footnote{The code is available at https://bitbucket.org/pandante/pyfc/src/master/} \citep{Lewis2002} which generates lognormal ``fractal cubes''. Within this model the gas density fluctuations in the ambient medium have the lognormal distribution and a Kolmogorov power-law spectrum with spatial index $\beta=5/3$. {The density field has the mean $\langle \rho \rangle$ and the standard deviation $\sigma$.} {Regardless the density variations we assume pressure equilibrium at the initial state, i.e., $\rho T={\rm const}$, with zero gas and dust velocities.} Among others the fractal distribution is characterized by a lower cutoff wavenumber $k_{min}$. Varying this value we restrict sampling to scales shorter than $N_c/k_{min}$, where $N_c$ is the number of cells along the direction of the fractal cube. It controls the largest size of fluctuations. In our models the dispersion { varies} from 0.2 to 3.0, so that the density contrast related to the mean value reaches up to 300 for the highest $\sigma$. The cuttoff wavenumber is taken from 4 to 20, that corresponds to the {longest length of fluctuations from 25 to 5~pc, respectively, for { $N_c = 256$ within the domain of $(96~\rm{pc})^3$ corresponding to spatial resolution 0.375 pc. This is} sufficient for { resolving the thermal (cooling) length estimated as $\lambda_t \sim 5~n^{-1} T_6$~pc and then for} adequate treatment of dynamics of a SN bubble in a clumpy medium (see a note about cell size for different density of an ambient gas below).

\begin{table}
\caption{The list of models.}
\center
\begin{tabular}{ccccc}
\hline
\hline
{\bf Model}   & {\bf Density}      & {\bf $k_{min}$} & {\bf $\sigma$ } & {\bf Dust/Sizes($\mu$m)/Bins }  \\
\hline
{\it  hp0  }  &  homogeneous &   --      & --         & poly / 0.003--0.3 / 11  \\
{\it  lp1  }  &  lognormal   &   16      & 0.2        & poly / 0.003--0.3 / 11  \\
{\it  lp2  }  &  lognormal   &   16      & 0.8        & poly / 0.003--0.3 / 11  \\
{\it  lp3  }  &  lognormal   &   16      & 1.5        & poly / 0.003--0.3 / 11  \\
{\it  lp4  }  &  lognormal   &   16      & 2.2        & poly / 0.003--0.3 / 11  \\
{\it  lp5  }  &  lognormal   &   16      & 3.0        & poly / 0.003--0.3 / 11  \\
\hline
{\it  hm0  }    &  homogeneous &   --      & --         & mono / 0.1 / 1  \\
{\it  lm1  }    &  lognormal   &   16      & 0.2        & mono / 0.1 / 1   \\
{\it  lm2  }    &  lognormal   &   16      & 0.8        & mono / 0.1 / 1   \\
{\it  lm3  }    &  lognormal   &   16      & 1.5        & mono / 0.1 / 1   \\
{\it  lm4  }    &  lognormal   &   16      & 2.2        & mono / 0.1 / 1   \\
{\it  lm5  }    &  lognormal   &   16      & 3.0        & mono / 0.1 / 1   \\
\hline
{\it  lm6  }    &  lognormal   &   4       & 2.2        & mono / 0.1 / 1   \\
{\it  lm7  }    &  lognormal   &   8       & 2.2        & mono / 0.1 / 1   \\
{\it  lm8  }    &  lognormal   &   12      & 2.2        & mono / 0.1 / 1   \\
{\it  lm9  }    &  lognormal   &   20      & 2.2        & mono / 0.1 / 1   \\
\hline
\hline
\end{tabular}%
\label{tab-models}
\end{table}

{ In the initial state the dust follows the gas density distribution} with the dust-to-gas ratio equal to $0.01$, which is usual for a solar metallicity value adopted in our models. At the initial moment we { consider polydisperse dust grains with the MRN size spectrum \citep{Mathis1977} with size range $0.003-0.3\mu$m in} $N_{s}=11$ binned equally in the log-scale, and (b) single-size (or {\it initially} monodisperse) grains of $a_0 = 0.1\mu$m. { In the latter case the monodisperse size distribution evolves gradually to a polydisperse one due to sputtering and emergence of small size particles.} In our simulations we assume the minimum grain radius of 0.001$\mu$m. The parameters of several models considered here are summarized in the Table~\ref{tab-models}. 

We inject the mass and energy of a SN in a region of radius $r_0=1.5$~pc, assuming commonly used values $30~\msun$ and $10^{51}$~erg. The energy is injected in thermal form. The injected mass of metals is $10 ~\msun$. Here we do not consider the dynamics of dust grains injected by SN, we will consider it elsewhere. Several aspects related to their evolution can be found in \citet{vs2023}.

\begin{figure*}
\center
\includegraphics[width=15.8cm]{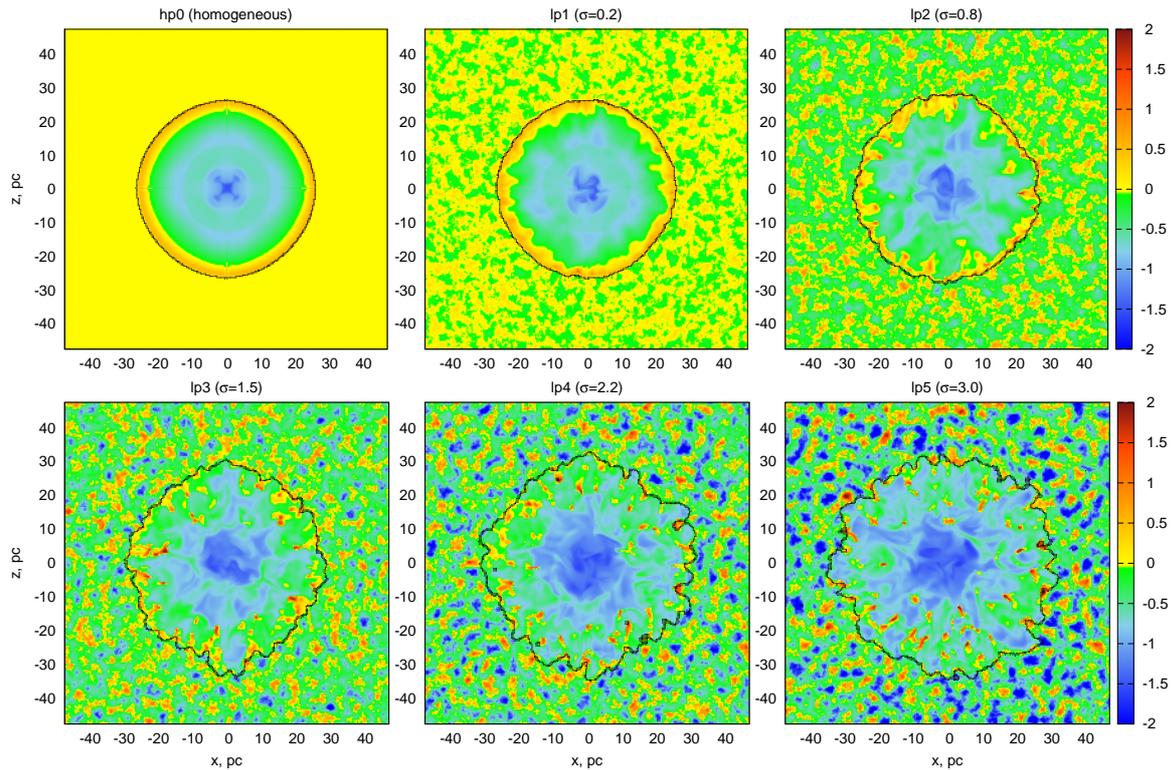}
\caption{
The 2D slices of gas number density (log[$n$, cm$^{-3}$]) at 50~kyr after SN explosion in a homogeneous medium in the model {\it hp0} (the upper left panel) and a clumpy medium in the models {\it lp1..lp5} (the other panels from left to right in the upper row and then in the lower row), i.e  for the following values of $\sigma = 0.2,\ 0.8,\ 1.5,\ 2.2$ and 3.0 with a fixed $k_{min}=16$. The black line shows the outer border of the SN bubble determined by the velocity jump at the shock front. 
}
\label{fig-den-maps}
\end{figure*}

Simulations are run consistently with tabulated non-equilibrium cooling rates fitting the calculated ones for a gas that cools isochorically from $10^8$ down to 10~K \citep{v11,v13}. The heating rate is assumed to be equal to a value chosen such as to stabilize the radiative cooling of the ambient gas with pressure $nT \simeq 10^4$cm$^{-3}$K. The mean value of the gas density is set $\rho_0=\langle \rho \rangle=1$~cm$^{-3}$ as fiducial. The total mass in the computational domain is the same in all models. We have run a set of models with mean densities equal to 0.3, 3 and 10cm$^{-3}$ for a clumpy medium with parameters for model '{\it lp4}', i.e. $\sigma=2.2$ and $k_{min}=16$. In these runs the spatial resolution is set to 0.5, 0.375 and 0.1875~pc, respectively.

We use our gasdynamic code \citep{vns2015,vsn2017} based on the unsplit total variation diminishing (TVD) approach that provides high-resolution capturing of shocks and prevents unphysical oscillations, and the Monotonic Upstream-Centered Scheme for Conservation Laws (MUSCL)-Hancock scheme with the Haarten-Lax-van Leer-Contact (HLLC) method \citep[see e.g.][]{Toro2009} as approximate Riemann solver. This code has successfully passed the whole set of tests proposed in \cite{Klingenberg2007}. In order to follow the dynamics of dust particles we have implemented the method \citep[see description and tests in Appendix A of][]{vs2023} similar to that proposed by \citet{Youdin2007}, \citet{Mignone2019} and \citet{Moseley2023}. The backward reaction of dust on to gas due to momentum transfer, work done by the drag force and the frictional heating from dust particles are also accounted in order to ensure both dynamical and thermal self-consistency. We take into account the destruction of dust particles by both thermal (in a hot gas) and kinetic (due to a relative motion between gas and grains) sputtering \citep{Draine1979b}. 

The mass of dust { pre-existing} in a medium is redistributed between many dust 'superparticles'. Each 'superparticle' consists of  numerous physical grains of a single size. To follow the dust transport in a medium we set at least one 'superparticle'  per a computational cell. For polydisperse dust we adopt one 'superparticle' in each bin of the size distribution of grains. This results in the total number of { pre-existing} (in the surrounding ISM) dust 'superparticles' in the domain $256^3 N_{s} \sim 16 N_{s}$ millions, where $N_{s}$ is the number of dust size bins in the distribution. Each 'superparticle' contains the total mass of single-size dust particles inside a cell. So the total dust mass in a cell is a sum of masses of 'superparticles' inside a cell.


{ Sputtering is effective and dominant in a hot gas. In a radiatively cooling gas other destructive processes can come into play. }

At temperatures $T\simlt 10^5$~K dust particles can be destroyed via shattering in grain--grain collisions provided the number of small size particles is considerable and their relative velocities are high \citep{Jones1996,Hirashita2009,Murga2019}. { Let us estimate the mass fraction of grains, whose velocity dispersion is necessary for efficient shattering $\sigma_v \simgt 30$~km~s$^{-1}$ \citep[e.g.,][]{Murga2019}. In addition, we consider grains located in a gas with $T\simlt 10^5$~K. Due to the latter condition the mass fraction sharply grows to $\sim0.2$ at the age $\sim 70$~kyr in the bubble expanding in a homogeneous medium. 
In an inhomogeneous medium this fraction starts to grow earlier, but reaches lower values, e.g. in a clumpy medium with density dispersion $\sigma=2.2$ it is about $\sim 0.04$ at $\simgt 20$~kyr. This represents the estimates of the dust mass contained under conditions favourable for shattering, but this does not mean that such grains remain under such conditions within sufficient time period.} { Several estimates give that this process becomes important on timescales longer than $\sim 1-5$ Myr in the warm ionized medium with $(T,n)=(8\times10^3$K,~0.1cm$^{-3})$ and longer than several tens of Myr in warm $(6\times10^3$K,~0.3cm$^{-3})$ and cold $(10^2$K,~30cm$^{-3})$ neutral medium \citep{Hirashita2009}}.
For the conditions in SN remnants the timescale for shattering { ranges} in $\sim 4-40$Myr \citep{Martinez2019}. Applying the post-processing calculations to the 3D SN evolution, \citet{Kirchschlager2022} have found that grain-grain collisions can be { important} in destroying { pre-existing dust by} a SN bubble older more than several hundred thousand years. { These scales are much longer than the final time of our runs. However, the estimates are seen to be model dependent. The question of dust metamorphosis under grain-grain collisions deserves a separate study.}

{ 
In the presence of magnetic fields the grain--grain collisions are believed to enhance their role in the dust destruction \citep[see e.g.][and references therein]{McKee1987,Seab1987}. 
Due to betatron acceleration the larger grains reach higher velocities, therefore they experience more destruction. Therefore, neglecting grain--grain collisions due to magnetic fields can lead to underestimation of the overall dust destruction rate. However, this depends on cooling efficiency and on inhomogeneity of a medium. 
Moreover, a rapid decrease of pressure behind the shock front leads not only to reduction of thermal sputtering, but also to the betatron deceleration of grains in the postshock magnetic field \citep{McKee1987}. In these conditions statistical properties of clumpy density field should play a significant role. For instance, one can note that in turbulent magnetized flows an enhanced gas-grain coupling from the Lorentz force provides to grains relative protection from shattering \citep{Moseley2023}. These studies demonstrate a very complex behavior of grains in magnetic field, and show that the efficiency of grain-grain collisions deserves a further consideration, particularly when it concerns a clumpy magnetized ISM.  
}




\section{Results}

\subsection{Shock penetration through the clumpy medium}

Fig.~\ref{fig-den-maps} illustrates propagation of the shock from a SN explosion in a medium with different characteristics of density distribution at the SN age $t=50$ kyr. The black line shows the outer border of the SN bubble determined by the velocity jump at the shock front. At first, one can note an increase of inhomogeneities inside the bubble for higher $\sigma$. Fragments located very close to the origin of the shock wave have been completely disintegrated, their gas replenishes the interior of the bubble. Fragments positioned further from the origin are destroyed only partly, they { survive} behind shock wave even in a surrounding medium with moderate density dispersion in the model {\it lp3} with $\sigma = 1.5$. They { are} not been supported further { on} by the external pressure, thus, they expand and replenish the hot rarefied interior of the bubble by warm denser gas that { contains} unprocessed dust grains. Therefore, the destruction of fragments leads to increase of radiative losses.

\begin{figure}
\center
\includegraphics[width=8.5cm]{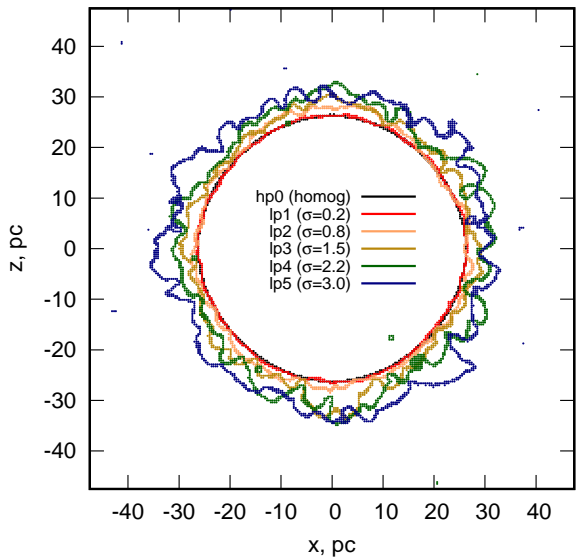}
\caption{
The outer border of the SN bubble determined by the velocity jump at the shock front in the models {\it hp0} and {\it lp1..lp5} at $t=50$kyr. These borders are shown by black lines in Fig.~\ref{fig-den-maps}.
}
\label{fig-borders}
\end{figure}

At second, a higher density contrast of inhomogeneities results in a faster penetration of the shock wave through the clumpy ISM as seen in Fig.~\ref{fig-borders}. This is because the shock wave encountering a denser region flows around it, penetrate into the low-density intercloud gas and restores the front in ambient gas after passing a denser obstacle with higher velocity $v\propto \rho^{-1/2}$ \citep[e.g.][]{Korolev2015,Slavin2017,Wang2018}. In { a} more inhomogeneous medium (higher $\sigma$) the shock wave between clumps remains adiabatic for a longer period of time and propagates through diffuse low-density gas with higher velocity.

\begin{figure}[h]
\center
\includegraphics[width=8cm]{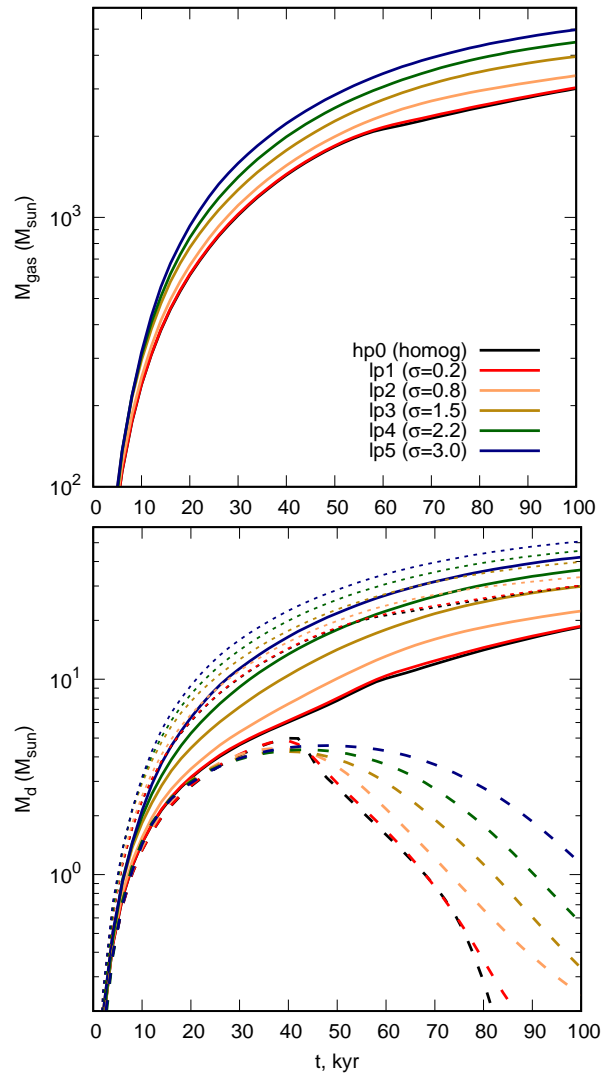}
\caption{
Evolution of gas (upper panel) and dust (lower panel) mass inside the SN bubble for different models: {\it hp0} and {\it lp1..lp5} as indicated in the legend. Solid lines show evolution { with} sputtering, for dotted lines sputtering is { not} accounted; dashed lines in lower panel depict the dust mass included in hot gas ($T>10^6$K).
}
\label{fig-mass-evol}
\end{figure}

\begin{figure}[h]
\center
\includegraphics[width=8cm]{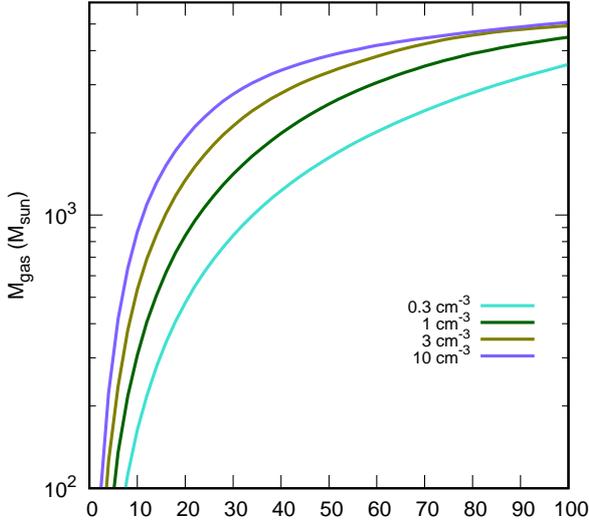}
\caption{
The same as in the upper panel of Fig.~\ref{fig-mass-evol}, but for the model {\it 'lp4'} (a clumpy medium with $\sigma=2.2$) with mean densities equal to 0.3, 1, 3 and 10cm$^{-3}$.
}
\label{fig-mass-evol-n}
\end{figure}

Due to such propagation of the SN shock the gaseous mass inside the bubble increases during evolution in more inhomogeneous medium with higher $\sigma$ (upper panel Fig.~\ref{fig-mass-evol}). The mass swept up in the medium with { a} low amplitude of inhomogeneity ($\sigma \simlt 0.2$, { as in} models {\it hp0} and {\it lp1}) after the free expansion phase at a few kyr evolves as $\sim t^{6/5}$ untill $\sim 30-40$~kyr which corresponds to the adiabatic phase, and as $\sim t^{3/4}$ afterwards when radiative losses come into play.  Increase of the amplitude of density inhomogeneity leads to a mixing of phases in different parts of the shell due to their interaction with ambient gas of different density. In an intercloud medium of lower density both transitions between the expansion regimes, from free expansion to adiabatic phase and then to radiative one, occur later, the shock moves with higher velocity and engulfs larger volume including dense clumps. As a consequence, the gas mass inside the bubble increases faster with the inhomogeneity magnitude. The dust mass inside the bubble also grows (lower panel Fig.~\ref{fig-mass-evol}). Dust particles { pre-existing} in the ambient medium are swept up together with gas and cross the shock front. Depending on the local shock velocity of the shell the conditions behind the front can be favourable for sputtering of grains. This results in a decrease of the total dust mass as seen from comparing the solid { (with sputtering)} and dotted { (without sputtering)} lines in Fig.~\ref{fig-mass-evol} (lower panel). In Fig.~\ref{fig-mass-evol} only results for polydisperse dust are presented. Monodisperse dust demonstrates  similar behaviour, as can be seen below.

In case of different mean density of the ambient medium $\langle \rho \rangle=0.3$ and $3$~cm$^{-3}$ the total mass of gas in the bubble follows similar power-law dependences as seen in Fig.~\ref{fig-mass-evol-n}: adiabatic ($\sim t^{6/5}$) and radiative ($\sim t^{3/4}$) phases. Only for $\langle \rho \rangle\simgt 3$~cm$^{-3}$ the bubble expansion is decelerated significantly after $\simlt 80$kyr { because the} flow becomes highly mass-loaded, and the total mass grows slower than $\sim t^{0.4}$ \citep[see also][]{Korolev2015}. For $\langle \rho \rangle = 10$~cm$^{-3}$ the bubble expansion becomes rather slow after $\sim 40$kyr. 

{ It is important to emphasize that the influence of various dust particles} on the SN bubble dynamics is negligible, thus, {the evolution of the bubble} is almost the same in the models both with initially monodisperse ({\it 'hm0'}, {\it 'lm1..lm5'}) and polydiperse ({\it 'hp0'}, {\it 'lp1..lp5'}) dust. Moreover, the spatial size of clumps does not { also} remarkably influence on the global inflow of the dust mass through the (whole) surface of the SN bubble in the model {\it 'lm4'} and the set of the models {\it 'lm6..lm9'} with the maximum spatial size varying from 25 to 5pc for the fixed dispersion $\sigma=2.2$. We expect the same for the models with polydiperse dust and consider below the models with varying density dispersion value $\sigma$ and focus on the models with polydisperse dust {\it 'hp0', 'lp1..lp5'} (Tab.~\ref{tab-models}).

\subsection{Dust destruction in the clumpy medium} 

The upper panel of Fig.~\ref{fig-mass-surf} presents the evolution of the 'dust mass survival' fraction $f_m=M_d/M_{d0}$, i.e. the ratio of the total mass {of dust grains} inside the SN bubble (behind { the} SN shock) normalized to the dust mass without taking into account grain destruction $M_{d0}$. {The evolution of both masses in models with polydisperse dust {\it hp0} and {\it lp1..lp5} { is} depicted in lower panel of Fig.~\ref{fig-mass-evol}}. { As said above,} only { pre-existing} dust is considered, without accounting the injected by the SN.

\begin{figure}
\center
\includegraphics[width=8.5cm]{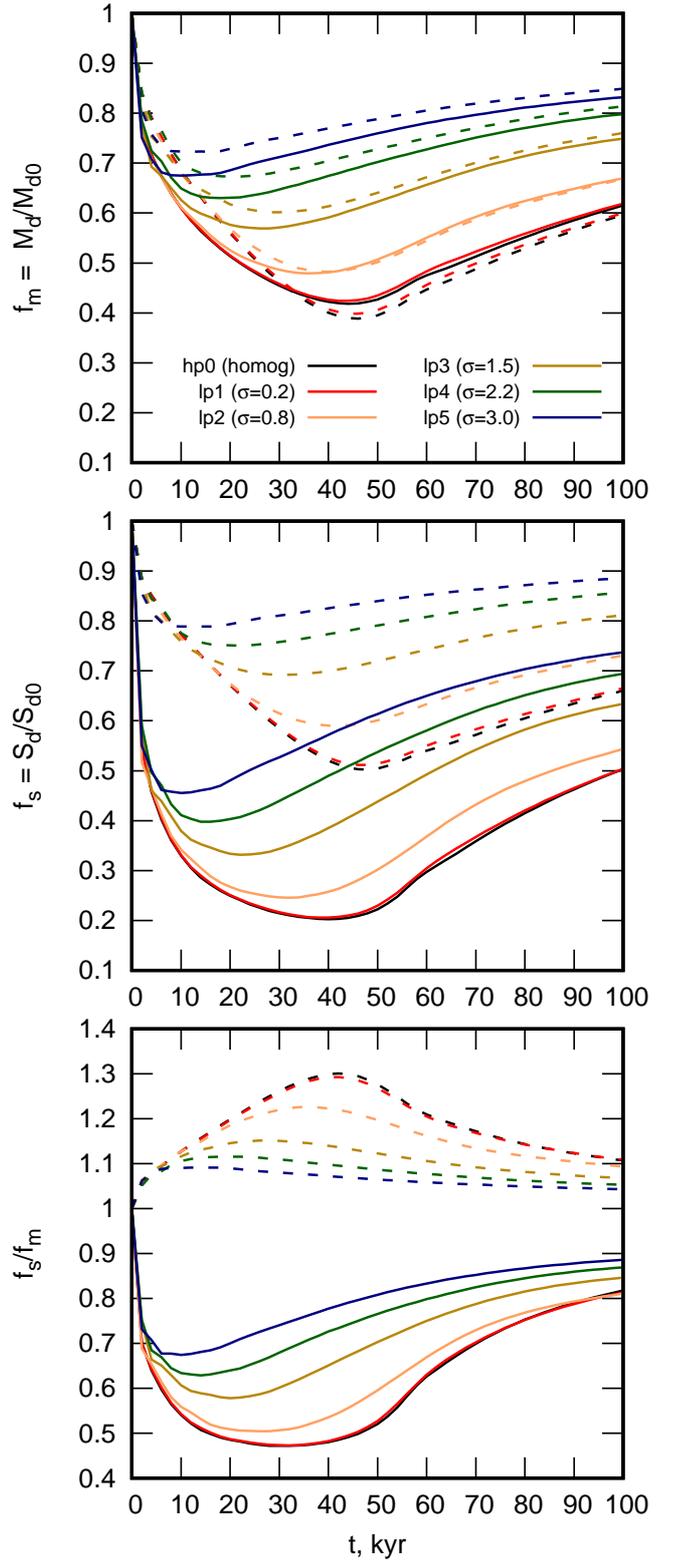}
\caption{
Total mass and cumulative surface of the { destroyed} dust grains inside the SN bubble (behind SN shock) normalized to the values without  accounting grain destruction, $M_{d0}$ and $S_{d0}$: $f_m = (M_d/M_{d0})$ (upper panel) and $f_s = (S_d/S_{d0})$ (middle panel), and their ratio $f_s/f_m = (S_d/S_{d0})/(M_d/M_{d0})$ (lower panel). Solid and dashed lines  correspond to the models for polydisperse and initially monodisperse dust, respectively. Black and color lines present the evolution in homogeneous and clumpy media as show in the legend.
}
\label{fig-mass-surf}
\end{figure}

In all models with polydisperse dust (solid lines in upper panel of Fig.~\ref{fig-mass-surf}) there is a significant drop of the value $f_m$ during first $1-2$kyr, when the shock wave heats a gas up to $T\sim 10^7$K. For a homogeneous medium (the model {\it 'hp0'}) this period corresponds to the transition from free expansion to adiabatic phase. At the age of $t\sim 2-3$kyr one can note a small difference between the evolution in homogeneous and clumpy { media. However, at $t \simgt 10$~kyr the difference increases and becomes more pronounced (upper panel of Fig.~\ref{fig-mass-surf}). The survived dust fraction $f_m$ decreases until the SN shock velocity drops below a certain threshold depending on $\sigma$. Before this threshold is reached} the fraction $f_m$ goes on down: the { pre-existing} dust crossed the shock front is efficiently sputtered because the bubble's interior is still hot, especially for low-clumpy medium with $\sigma \simlt 1.5$ (the models {\it 'lp1..lp3'}). 

{ In more inhomogeneous media (the models {\it 'lp4..lp5'}) radiative losses inside the bubble are higher because of the mass load from destroyed clouds located in the innermost of the SN bubble.} This leads to earlier beginning of the radiative { phase followed by a deceleration of the most massive part of the shell and a decrease} of gas temperature inside the bubble. { This} unavoidably results in suppression of dust destruction. The bubble continues expanding and sweeping the unprocessed { pre-existing} dust. Consequently, the 'dust survival' fraction { turns} to grow. The turnaround point { with} the minimum value $f_m$ corresponds to the end of efficient destruction of grains. After this time the dust mass { in} the bubble increases due to sweeping up the { pre-existing} grains. In a homogeneous medium this occurs { immediately} after the shell has become strongly radiative, $t\sim 40$kyr (see dash black line in upper panel of Fig.~\ref{fig-mass-surf}). In a clumpy medium { with} $\sigma=2.2$ (model '{\it lp4}') this takes place earlier: after a period of 10kyr the fraction $f_m$ starts to grow and follows the evolution of the gas mass at the radiative phase as $M\sim r^{-3/4}$.

Thus, the clumpiness of the medium directly affects the survivability of the dust grains against sputtering: the higher { the} dispersion of density the larger amount of dust survives. It primarily comes from the fact that the shock front engulfs a larger amount of gas and dust during its propagation through the more inhomogeneous medium (Fig.~\ref{fig-mass-evol}). A major part of dust is contained into dense fragments, where the shock wave penetrates with lower velocity and cannot heat a gas enough to { make the} sputtering efficient. One can see that the mass fraction of the survived dust grows { about a factor of 1.7 in models {\it 'hp0'} and {\it 'lp5'}. This value comes from the comparison of the mass fractions at the age of $\sim 40$ and $\sim 10$kyr in these models, respectively,}  when the radiative phase starts. During further evolution the shell sweeps up unprocessed { pre-existing} dust, and the difference between models {\it 'hp0'} and {\it 'lp5'} decreases to a factor about 1.3 at $t=100$kyr -- the end of our calculations. 

Besides, we consider the evolution of the SN bubble in a medium with { pre-existing} monodisperse dust particles. In these models the 'dust survival' fraction shows similar behaviour (see models {\it 'hm0'} and {\it 'lm4..lm5'}, dashed lines in upper panel of Fig.~\ref{fig-mass-surf}). A { slightly slower} decrease { at early times} compared to the models with polydisperse dust is explained by the {absence} of small particles $a \simlt 0.01\mu$m, which are efficiently sputtered { in} this period. 

The 'dust mass survival' fraction demonstrates no significant difference between models with initially mono- and polydisperse grains.  In case of smaller initial size of monodisperse dust compared to $a_0=0.1\mu$m, { which is taken in our models,} the difference between models {\it 'hm0'} and {\it 'lm5'} becomes { higher}, but only until the beginning of radiative phase. Later, dust swept up becomes dominant by mass and this difference tends to be close to a factor about 1.3 at $t=100$kyr.

In polydisperse dust small-size particles normally dominate by number. Moreover, for { a steep power-law} dust size  distributions with $n(a)\propto a^{-p}$ like MRN, small particles contribute predominantly in extinction. In addition, smaller dust particles are usually hotter -- in equilibrium { their temperature} varies approximately as $T_d\propto a^{-1/6}$, and as such they determine dust emission  at higher frequencies. In order to illustrate this we present here how total surface of grains is changed during the SN propagation in a cloudy medium. 

Middle panel of Fig.~\ref{fig-mass-surf} presents the cumulative surface\footnote{Note that in our simulation we describe the dynamics of dust in terms of 'superparticles'. Each 'superparticle' consists of numerous identical microparticles or grains. It is easy to calculate the number of grains in a 'superparticle' and their cumulative surface.} of the destroyed dust grains behind the SN shock front normalized to the value without taking into account grain destruction $f_s = (S_d/S_{d0})$, { hereafter the} 'dust surface survival' fraction. For polydisperse dust, small-size grains of $a\simlt 0.01\mu$n are more efficiently destroyed in hot gas during { the} first 10kyr. For { a MRN dust size distribution} such grains contribute to the total surface larger than to the total mass, thus, the fraction $f_s$ drops faster and deeper than the value $f_m$. The value of $f_s$ decreases by a factor of $\sim 3$ for SN explosion in a homogeneous medium (solid black line, model {\it 'hp0'}). The hostile conditions for small-size grains are supported in the bubble before it goes to radiative phase at $t\sim 30-40$kyr. Until this time the cumulative surface of grains decreases by almost 5 times. Inhomogeneity in the ambient gas leads to significant inhibition of dust destruction (see models {\it 'lp1..lp5'}), thus, the total surface of dust increases.

At later stages, however, the total surface area of the dust particles experiences an increase, provided by both less efficient sputtering behind decelerated shock in dense clumps and sweeping of unprocessed dust by the expanding SN shell.
{ Till 100~kyr the 'dust surface survival' fraction increases two times compared to its minimum value reached at the moment, when the radiative phase in the SN remnant starts.} 

In models with initially monodisperse dust the evolution of the mass and surface fractions, $f_m$ and $f_s$, presented by dashed lines in upper and middle panels of Fig.~\ref{fig-mass-surf}, decrease until the radiative phase begins. { Their behaviour depends on the transition to the radiative phase and is similar to that in models with polydisperse dust described above.}

The difference between the evolution of the mass and surface 'survival' factors is more clear observed by using their ratio presented in lower panel of Fig.~\ref{fig-mass-surf}. For polydisperse dust, the ratio $f_s/f_m$  is determined by sputtering of small-size grains. The cumulative surface depends on grain size as $\sim a^{-0.5}$ for the MRN distribution, whereas the mass $M(a) \sim a^{0.5}$. Hence, the destruction of small-size grains leads to a remarkable drop of the ratio $f_s/f_m$ during first 10kyr (as shown by color lines, models {\it 'lp1..lp5'}). At $t\simlt 30$kyr, the shock velocity remains higher 100km/s for the evolution in an ambient gas with low density dispersion. The period with low ratio $f_s/f_m$ continues until entering the radiation phase. This period is shorter for a  higher value of $\sigma$ and for a higher mean gas density. For $\sigma=0$ it has completed around $40-50$kyr (solid black line, model {\it 'hp0'}), whereas for $\sigma=2.2$ it has finished already at $\sim20$kyr (solid red line, model {\it 'lp4'}).

For monodisperse dust the cumulative surface depends on grain size as $\sim a^2$, while the total mass $\sim a^3$, the value $f_s/f_m$ is proportional to $\sim 1/a(t) \sim \dot a t$, before sputtering remains efficient (for $T>10^6$K the sputtering rate $\dot a$ weakly depends on gas temperature and is about $\sim 10^{-3} n~\mu$m/kyr, see eq. 25.14 in \citet{Draine-book}). This corresponds to the linear part of the curve for the evolution in a homogeneous medium (dash black line, model {\it 'hm0'}). Increase of inhomogeneity in the ambient gas leads to earlier breaking of this dependence. At late { times} the ratio obviously tends to 1 due to growth of the fraction of unprocessed dust in the total mass of dust inside the bubble.

\begin{figure}
\center
\includegraphics[width=8.0cm]{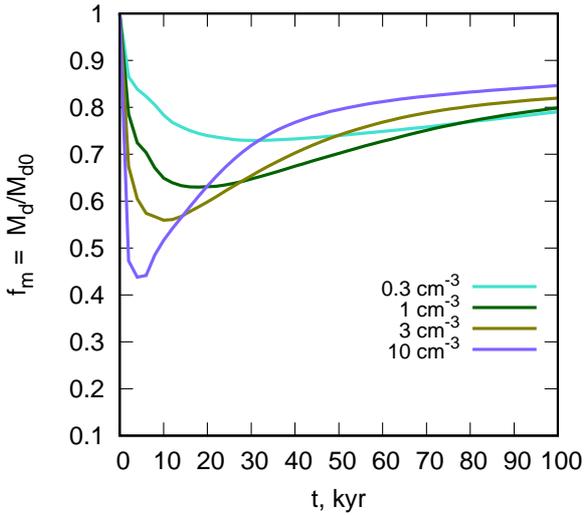}
\caption{
The same as in the upper panel of Fig.~\ref{fig-mass-surf}, but for the model {\it 'lp4'} (a clumpy medium with $\sigma=2.2$) with mean densities equal to 0.3, 1, 3 and 10cm$^{-3}$.
}
\label{fig-mass-surf-n}
\end{figure}

The overall trend in evolution of the survival dust mass fraction $f_m$ is determined by interrelation between radiative gas cooling and dust sputtering in different environment. In conditions with higher $\langle\rho\rangle$ { a more efficient dust sputtering at early stages weakens rapidly as the SN shock enters the radiation dominated expansion. As a result, the sputtering exhausts} and $f_m$ turns growing. In conditions with lower $\langle\rho\rangle$ the post-shock gas stays hot longer and sputtering $\propto\langle\rho\rangle$ is less strong, { then, $f_m$ decreases slower and starts to grow after transition to radiative phase} as seen in Fig.~\ref{fig-mass-surf-n}. For the fiducial value of density 1~cm$^{-3}$ the 'dust mass survival' fraction $f_m=M_d/M_{d0}$ reaches minimum $\sim 0.6$ at the SN age of $\sim 15$kyr. { For} higher mean density it becomes lower and shifts to earlier time, e.g. it is $\sim 0.4-0.5$ at $\sim 4$~kyr for  $\langle \rho \rangle = 10$~cm$^{-3}$ (Fig.~\ref{fig-mass-surf-n}). Later, more mass of dust is swept up and the fraction $f_m$ grows as $\sim t^{0.3}$ (Fig.~\ref{fig-mass-surf-n}). { At later stages mass-loading effects inhibit growth of $f_m$} to $\sim t^{0.1}$, so the value $f_m$ nearly saturates at $\sim 0.8$, that is close to $f_m$ for the fiducial mean density.

\subsection{Dust in various thermal phases} 
\label{sec-thermal}

\begin{figure*}
\center
\includegraphics[width=15.9cm]{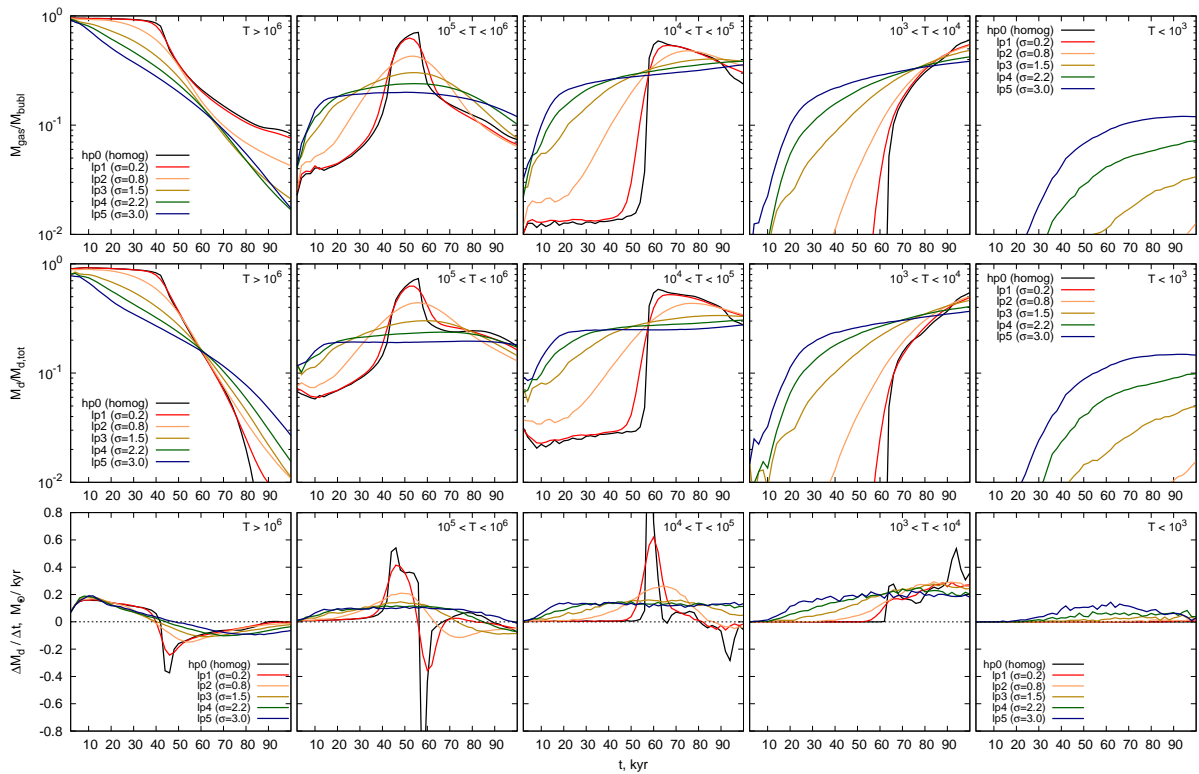}
\caption{
Total mass of gas (upper row) and dust (middle row) located in different thermal phases normalized to the total mass of gas and dust enclosed in the SN bubble in {models  {\it 'hp0'}, {\it 'lp1..lp5'} for polydisperse dust distributed within $0.003-0.3\mu$m by 11 bins. The growth rate of dust mass $\Delta M_d/\Delta t$ (in $\msun$/kyr) associated with various thermal phases (panels from left to right) of a gas in the SN bubble for the same models presented in Fig.~\ref{fig-m-t}. 
} 
}
\label{fig-m-t}
\end{figure*}

Fig.~\ref{fig-m-t} presents the distribution of the gas mass enclosed inside { different thermal phases of the bubble}: hot ($T>10^6$K), warm-hot ($10^5-10^6$K), diffuse ($10^4-10^5$K), warm-neutral ($10^3-10^4$K) and cold ($T<10^3$K) ones. During the evolution in a homogeneous medium (black line) the hot phase (left panel) is dominant as long as the bubble remains adiabatic, i.e until the age of $\sim 40$kyr. After that the gas goes to warm-hot and diffuse phases (see second and third panels). At $t\sim 60$kyr { the warm-neutral phase rapidly grows and reaches more than 20\% already at $t\simgt 70$~kyr (fourth panel).}  This gas is { confined} in the dense SN shell. During the calculation time no gas cools down below $10^3$K (fifth panel). It is { seen}, there is a clear sequence of thermal phases during the bubble evolution: each phase dominates in its epoch, they do not mix each other except rather short periods.

In a cloudy medium the radiative phase starts earlier in {those} parts of the shell, where it interacts with the dense clumps. It begins almost immediately after the explosion { because of an enhanced cooling in dense cloudlets} close to the SN origin. As a consequence, the fraction of hot gas decreases gradually since this early time (colored lines in left panel): the fraction becomes lower for higher $\sigma$ because of higher volume filling factor of denser clumps. This results in an earlier emergence of warm-hot, diffuse, warm-neutral gas (see second to fourth panels). For $\sigma\simgt 1.5$ a remarkable mass fraction can be found in the cold phase (fifth panel). The warm-neutral and cold phases are located inside those several clumps which have not been completely { destroyed} by the SN shock and have remained far behind the shock front in the hot bubble due to inefficient acceleration (see Fig.~\ref{fig-den-maps}). 

At first glance, dust { is associated with} the thermal phases in comparable proportions as the gas itself (see middle row in Fig.~\ref{fig-m-t}). However, one can note several differences. The most remarkable one is found for the hot phase, where the interrelation between the dust and gas fractions changes at $t\sim 60$kyr (close to the intersection point of all lines in left middle panels of Fig.~\ref{fig-m-t}).

Let us consider the evolution in a homogeneous medium (model {\it 'hp0'}). Interstellar grains associated with hot gas are swept up by the SN shock, while it was strong enough to transfer the gas to the hot phase. A part of such particles has penetrated into low-dense gas behind the front \citep[see e.g. Fig.~1 in][]{vs2023}. { This} occurs while the remnant is young and the shell moves with higher velocity. Within this period, extending to $\sim 40$kyr, both gas and dust fractions in the hot phase are close to unity ({see black lines depicted the model {\it 'hp0'} in the} left upper and middle panels in Fig.~\ref{fig-m-t}).  

During further evolution the shell decelerates and cools down, such that the dust contained there is associated with gas of lower  temperature: one can note { a} rapid growth of the dust and gas in warm-hot phase (second column of panels in Fig.~\ref{fig-m-t}). Dust particles penetrated deeper get higher momentum from the high-velocity gas. Afterwards their velocity remain high because they move through hot rarefied gas behind the shell. When the bubble enters the radiative phase the shell decelerates, grains located far behind the shell still move with higher velocity. Therefore, such grains overcome the shell \citep[][]{Slavin2020,vs2023}, which becomes cooler by that time. Thus, the grains become to be associated with lower thermal phase. { After the age of} $\sim50-60$kyr,  the hot gas { becomes almost free of dust within $10-20$~kyr}: the dust fraction { associated with hot gas decreases} from $\sim0.1$ at $\sim 60$kyr to $\simlt 0.01$ at $\simgt 80$kyr, while the { fraction of hot gas reaches} $\sim 0.1$ at 100~kyr.

The transitions of dust mass between various thermal phases of the gas can be clearly seen on the growth rate of dust mass {enclosed in the bubble} $\Delta M_d/\Delta t$ (lower row of Fig.~\ref{fig-m-t}). { In} a homogeneous medium (black line) the dust mass in the hot phase { grows at a} rate of $\sim 0.1\msun$/kyr until the age of $\sim 40$kyr. Afterwards the radiative phase begins, the rate drops dramatically and becomes negative (left panel), i.e. the dust leaves hot gas and goes to warm-hot gas. The rate for the warm-hot phase increases { sharply} around $\sim 40$kyr (second lower panel). { Later on a similar transfer of dust from hotter to colder phases can be observed in third to fifth panels in the lower row.}

{ In} an inhomogeneous medium { the} transition of the SN shell from adiabatic phase to radiative one { is less pronounced}. The thermal { structure} of the shell is mixed: when the front moves through dense clouds the transition { occurs} much earlier, { when} the front { penetrates from low-density intercloud regions into clumps, and the shocked clump gas becomes radiative, whereas the poost-shock intercloud medium remains adiabatic.} Clouds located closer to the SN origin are destroyed completely, { though more} distant ones can survive during their interaction with the shock and the hot interior of the bubble. Thus, the gas that crossed the shock front replenishes not only the hot phase (as in a homogeneous medium), but also other phases: the material from { destroyed} clouds contributes to the warm-hot and diffuse phases, denser gas of clouds goes to the warm-neutral one. Such interchanges increase { with} $\sigma$. { As a result}, the mass fraction of hot gas decreases gradually with value of $\sigma$ (left upper panel of Fig.~\ref{fig-m-t}). The fractions of a gas of lower temperature{ s} evolve more smoothly for higher $\sigma$ (see the other panels in the upper row).

During the first $\sim 40$~kyr { an} increase rate of the dust mass in the hot phase weakly depends on $\sigma$ (left lower panel of Fig.~\ref{fig-m-t}). Any increase ceases at $\sim 40$kyr, when the radiative phase { begins at a given} averaged ambient density. In some parts of the shell expanding in the intercloud medium the cooling becomes efficient at { later} time. { These parts decelerate and dust confined in hot gas behind the shell catches up with} these parts later. Thus, for higher $\sigma$ the dust fraction follows closer to the fraction of hot gas (left upper and middle panels of Fig.~\ref{fig-m-t}). For $\sigma\simgt 0.8$ the dust masses enclosed in other thermal phases increase with { an} almost constant rate { because of the cooling enhanced by the gas incoming from clouds destroyed} behind the shock front (lower row of Fig.~\ref{fig-m-t}). 

{ Overall, the mass of dust incoming from the destroyed clouds into hot gas is higher than its depletion due to {\it in situ} sputtering.} For media with higher $\sigma$ this mass is larger as shown by dashed lines in the lower panel of Fig.~\ref{fig-mass-evol}). Thus, the increase of dust mass in the hot phase with increase of $\sigma$ (the models {\it 'lp1..lp5'}, left panel in middle row of Fig.~\ref{fig-m-t}) is determined by { grains confined initially in dense clumps and survived strong sputtering behind high velocity shocks in the low-density intercloud medium.}

Fig.~\ref{fig-mpdf-evol} presents the probability distribution functions (PDF) of { grain masses $f_m$} involved into the SN bubble at three moments; { at $t=0$ the MRN size distribution was adopted}. { In the course of sputtering larger grains are shifted towards the domain of smaller masses, as can be seen from} the decrease of their mass fraction  and the increase of deposit by smaller grains: the slope of the distribution becomes slightly flatter for later time. 

Fig.~\ref{fig-mpdf-t} present the PDFs of grains associated with the different thermal phases of a gas. { At} the age $t=50$kyr (upper left panel) the warm-hot phase dominates (green line), where { most fraction of dust mass is} contained (see Fig.~\ref{fig-m-t}). Only large grains of size $a\sim 0.1\mu$m { are} associated with the hot phase {(red line)}. At later time, dust { grains are redistributed nearly equally between the diffuse (blue) and warm-neutral (pink)}. Large grains of size $a\sim 0.1\mu$m can be found only in warm-hot gas. Other thermal phases (the hottest and coldest ones) are { almost lack of dust. In the hot phase this is} due to destruction by sputtering or transition to low-temperature phases. 

\begin{figure}[t]
\center
\includegraphics[width=8.5cm]{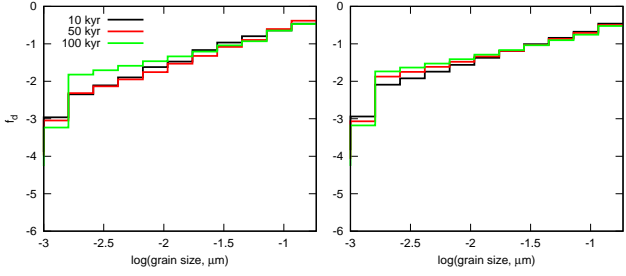}
\caption{
The mass PDF of grains involved into the SN bubble expanding in a homogeneous (left column, model {\it 'hp0'}) and clumpy (right column, model {\it 'lp4'}) at 10, 50 and 100kyr.
}
\label{fig-mpdf-evol}
\end{figure}

Such distribution can correspond to two populations of dust inside the bubble: { one} population represents  {\it `cold'} grains associated with the gaseous diffuse and neutral phases, the PDF of such grains is close to the initial one, the { other} consists of {\it `hot'} grains larger than 0.1$\mu$m and enclosed in the hot {(at 50~kyr, red line at upper left panel)} or warm-hot {(at 100~kyr, green line at lower left panel)} phases. {The mass} fraction of the second population is more than 10\%. Assuming cooling rate of a particle of radius $a$ as $L \sim n_e T^{1.5}$ in the thermal equilibrium \citep{Dwek1992}, where $n_e$ and $T$ are the electron number density and gas temperature, one can expect a remarkable influence on the total spectrum of dust. As the grain population associated with high-temperature gas is hotter, even a few percent fraction of such  dust can contribute considerably  to the dust spectrum \citep{Dwek1992}. More detailed calculations will be given in a companion paper elsewhere.

In a clumpy medium the PDFs demonstrate more regular structure: they have its own slope for each phase (see right column of panels in Fig.~\ref{fig-mpdf-t}). For low-temperature phases the slopes of the PDFs are close to that of the cumulative one {(black line)}. In high-temperature gas the distributions of grain sizes demonstrate steeper slopes due to efficient destruction of smaller grains. During the evolution gas cools down and dust associated with a given thermal phase is transferred to lower temperatures. Thus, the PDF slopes for dust enclosed in both hot and warm-hot phases become steeper for later time. The slope for the hot phase drops from almost $1$ at 50kyr to $2.5$ at 100kyr (see red lines, right panels in Fig.~\ref{fig-mpdf-t}). For the warm-hot one it changes from about $1$ at 50kyr to $1.5$ at 100kyr (see green lines, right panels in Fig.~\ref{fig-mpdf-t}). 

\begin{figure}
\center
\includegraphics[width=8.5cm]{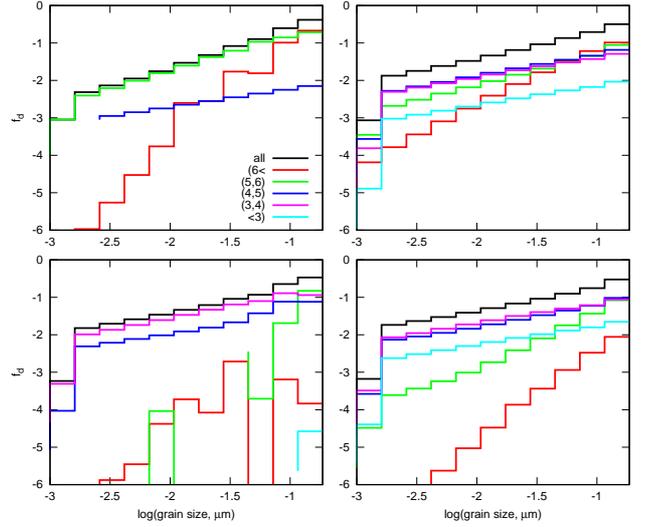}
\caption{
The mass PDF of grains located in a gas of various thermal phases at 50kyr (upper row) and 100kyr (lower row) for the SN bubble expanding in a homogeneous (left column, model {\it 'hp0'}) and clumpy (right column, model {\it 'lp4'}) medium. 
}
\label{fig-mpdf-t}
\end{figure}

\begin{figure}
\center
\includegraphics[width=8.5cm]{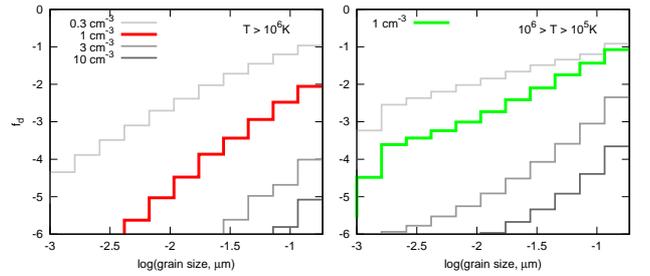}
\caption{
{
The mass PDF of grains enclosed in a gas of hot (left panel) and warm-hot (right panel) thermal phases as shown in the legend at 100kyr for the SN bubble expanding in a clumpy (right column, model {\it 'lp4'}) medium with the averaged gas number density $n=0.3, 1, 3, 10$cm$^{-3}$. The color lines depict the distributions for the fiducial value of density equal to $1$cm$^{-3}$ {as at lower right panel of Fig.~\ref{fig-mpdf-t}}.
}
}
\label{fig-mpdf-t-n}
\end{figure}

A change of {average} density of the ambient gas leads to corresponding variation of dust sputtering and modification of distribution over grain sizes (Fig.~\ref{fig-mpdf-t-n}). For particles associated with hot gas the PDF slope becomes flatter in the bubble evolved in low-density gas, e.g. for $\langle n \rangle = 0.3$~cm$^{-3}$ it equals to $1.5$ at 100kyr upper grey line in left panel of Fig.~\ref{fig-mpdf-t-n}). Increase of gas density related to the fiducial value of $1$cm$^{-3}$ facilitates highly efficient destruction of grains. The slope becomes only slightly steeper (almost $2$), but the mass fraction of grains as large as 0.1$\mu$m drops seriously: it decreases from 1\% for $1$cm$^{-3}$ to 0.01\% and 0.001\% for 3 and 10cm$^{-3}$, respectively, at 100kyr (left panel Fig.~\ref{fig-mpdf-t-n}). Therefore, one can conclude about almost absence of dust associated with the hot bubble evolved in the medium with $n\simgt 3$cm$^{-3}$.

For the warm-hot phase the fraction of dust in low-dense gas $n\sim 0.3$cm$^{-3}$ is around 10\% at $100$kyr (right panel Fig.~\ref{fig-mpdf-t-n}), it remains almost constant since $\sim 20$~kyr. The slope of the distribution is kept about $0.7$, which is close to the initial value equal to $0.5$. The warm-hot phase is more abundant by large particles, while the fraction of large grains in the hot phase of the bubble evolved in denser environments drops dramatically. For the fiducial gas density their fraction decreases negligibly. However, grains of smaller sizes are destroyed more efficiently. The slope becomes steeper with index about $1.5$ as mentioned above. Further increase of average ambient density leads to substantial decrease of large grains as $n^3$. This gives rather low fraction of large particles: about 1\% and 0.03\% for 3 and 10cm$^{-3}$, respectively. These values are small enough, but they are about 1.5--2 order of magnitude higher than those for the hot phase.

\section{Discussion and Conclusion}

Inhibited destruction of dust grains in a clumpy medium can help to facilitate a disbalance between dust production and destruction rates in the ISM of our and other galaxies. The ISM is well-known to be turbulent and clumpy within wide spatial scales. Thus, gaseous inhomogeneities shield dust grains from being destroyed by strong shock waves from SNe explosions: the total mass and cumulative surface of grains are higher in flows penetrating clumpy medium. This has several consequences that influence  radiation transfer across a SN bubble evolved in such medium, dust and gas emissivities and even molecule formation on grain surface. { These issues}  will be considered elsewhere.

We have studied how initially mono- ($a_0=0.1\mu$m) and polydisperse (the MRN distribution, $dn/da\sim a^{-3.5}$) dust grains { pre-existing} in an ambient medium are swept up by the SN shock wave and destroyed inside the SN bubble depending on magnitude of inhomogeneity (clumpiness), whose density field has the lognormal distribution with the mean $\langle \rho \rangle \sim 1$cm$^{-3}$ and the standard deviation $\sigma$, and a Kolmogorov power-law spectrum with spatial index $\beta=5/3$.

We have found the following:
\begin{itemize}
 \item Dust destruction inside the bubble is inhibited in more inhomogeneous medium: the larger amount of dust survives for the higher dispersion of density; the mass fraction of the dust for $\sigma = 2.2$ increases a factor of 1.7 in comparison with a homogeneous medium at the beginning of the radiative phase in the SN remnant. At the same SN ages the cumulative surface of grains due to inhibited destruction in a clumpy medium grows a factor of 2. Note that there is no significant difference of the mass fraction for models with initially mono- and polydisperse grains, whereas the evolution of cumulative surface depends significantly on shape of the grain size distribution;
 \item Inside the SN bubble older $\sim50$kyr evolving in a clumpy medium the mass fraction of dust associated with hot gas decreases slower than that in a nearly homogeneous medium. This is because the grains that cross the front earlier, move faster in a nearly homogeneous medium and during further expansion overtake the shell \citep[][]{Slavin2020,vs2023}, and leave the hot ($T\simgt 10^6$K) phase of a gas;
 \item Dust is getting efficiently destroyed in hot gas of the bubble expanding in a medium with low $\sigma$, and no significant dust fraction remains associated with hot gas. In more clumpy medium the income of unprocessed dust leads to flatter size distribution of dust enclosed in hot gas. 
 \item Inside the bubble evolving in a clumpy medium with higher mean density, dust is { destroyed} with higher rate, while { the} gas remains hot. However, this period is shorter and has completed earlier for denser medium. During further evolution, more mass of dust is swept up and the mass fraction of the survived dust grows and later saturates at level almost independent on gas mean density.
\end{itemize}

\section*{Acknowledgements}

We are thankful to Yuri A. Shchekinov for many valuable comments and Ilya Khrykin for his help. Numerical simulations of the bubble dynamics were supported by the Russian Science Foundation (project no. 23-22-00266).

\appendix
\section{Resolution test}
\label{sec-a}

{ 
We follow the evolution of an SN bubble with different spatial resolution for the grid. In our models the spatial resolution is set to 0.375~pc. At first we check a little lower resolution with cell size equal 0.5~pc and do not find significant differences in the  values averaged over the bubble as the dust 'survival' fractions and etc. Certainly, this is insufficient and we should check our approach for better resolution. However, the total number of particles is about 200 millions for our fiducial grid of 256$^3$ cells and number of bins equal to 11 in the size distribution for the models with initially polydisperse dust. Two times better resolution needs more than one billion particles. For initially monodisperse dust the requirements to the CPU time and memory for the dynamics of dust particles become weaker, but remain significant for the gas dynamics. Then, we should consider the resolution dependence for less time-consuming task.
}

{ 
We consider the interaction of a plane-parallel shock front with a clumpy medium. We inject energy in a several cells near the $xy-$boundary of the computation domain. The energy is added in thermal form. The energy density is $3\times 10^{-9}$~erg~cm$^{-3}$, this value is of the same order as inside the SN bubble of radius $\sim 15$~pc. Then, this plane-parallel shock can be considered as a part of the SN shell. At the inner $xy-$boundary we set the reflective condition. The shock front propagates along $z$-direction. The clumpy density field is generated using the same procedure as described in Sec.~\ref{sec-m}. Here the density dispersion $\sigma$ is taken equal to 2.2 and the cutoff wavenumber $k_{min}=16$. Then, the density distribution in this model is similar to that in model {\it 'hp4'}. At the initial moment we consider single-size (monodisperse) grains of $a_0 = 0.1\mu$m.
}

{ 
We take the computational domain of $(L_x\times L_y \times L_z)=(24\times 24\times 72~\rm{pc})^3$ with fiducial spatial resolution of 0.375 pc and increase this resolution two and four times: 0.18175~pc and 0.09375~pc. For instance, Fig.~\ref{fig-den-shock-maps} presents the 2D slices of gas number density for these cell sizes. One can note that the density distributions for different spatial resolution are similar each other behind the front. The position of the front is almost independent on resolution. Certainly, the thickness of the front decreases with increase of resolution and more fragments and other details can be found in the map for the highest resolution. However, the values averaged over large regions on the grid are expected to remain close for these runs.
}

{ 
We calculate the evolution of the dust mass and surface 'survival' fractions and their ratio behind the shock front. The same was done for the SN bubble expanding in a clumpy medium as depicted Fig.~\ref{fig-mass-surf}. Figure~\ref{fig-mass-surf-res} shows these values for different spatial resolution. One can note that the fractions demonstrate less than 5\% difference for models with the fiducial cell size $\Delta x = 0.375$~pc (label {\it 'x1'}) and four times smaller cell (label {\it 'x4'}). Then, we can conclude that the calculations with the fiducial spatial resolution describe adequately the dynamics of both gas and dust during the 'shock--clumpy medium' interaction and, as a consequence, during the SN bubble expansion in a clumpy medium.
}

\begin{figure}
\center
\includegraphics[width=7cm]{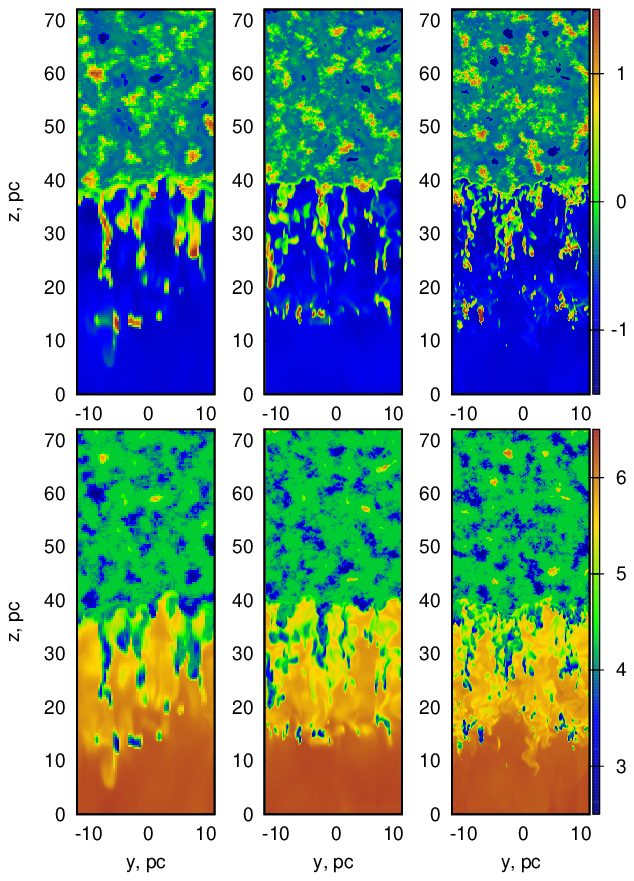}
\caption{
{ 
The 2D slices of gas number density (log[$n$, cm$^{-3}$], upper panels) and temperature (log[$T$, K], lower panels) at 150~kyr after starting interaction of shock front with a clumpy medium having the lognormal distribution of density with dispersion $\sigma=2.2$ and cutoff wavenumber $k_{min}=16$. The spatial resolution increases from standard cell size of 0.375~pc to two and four times better (the  panels from left to right).
}
}
\label{fig-den-shock-maps}
\end{figure}

\begin{figure}
\center
\includegraphics[width=8cm]{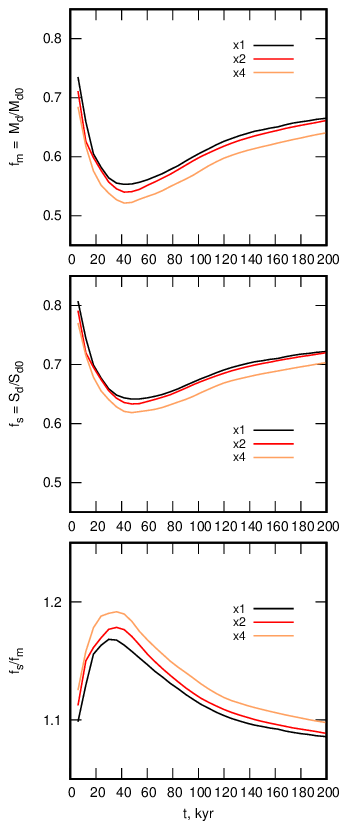}
\caption{
{ 
Total mass and cumulative surface of the destroyed dust grains behind the plane front normalized to the values without accounting grain destruction, $M_{d0}$ and $S_{d0}$: $f_m = (M_d/M_{d0})$ (upper panel) and $f_s = (S_d/S_{d0})$ (middle panel), and their ratio $f_s/f_m = (S_d/S_{d0})/(M_d/M_{d0})$ (lower panel). Black line presents the evolution in a clumpy medium with the standard spatial (labeled as $x1$) resolution. Color lines show the evolution for two ($x2$) and four ($x4$) times higher resolution.
}
}
\label{fig-mass-surf-res}
\end{figure}

\bibliographystyle{elsarticle-harv} 
\bibliography{p-bib1}

\end{document}